\documentclass{moriond}

\def\be{\begin{equation}}
\def\ee{\end{equation}}
\def\bea{\begin{eqnarray}}
\def\eea{\end{eqnarray}}

\newcommand{\Photo}{\includegraphics[height=25mm]{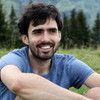}}

\begin{document}
\vspace*{4cm}
\title{FERMION MASSES AND MIXING FROM A MINIMUM PRINCIPLE}

\author{Rodrigo Alonso}

\address{Department of Physics, University of California at San Diego, 9500 Gilman Drive,\\ La Jolla, CA 92093-0319, USA}

\maketitle\abstracts{We analyze the structure of  quark and  lepton  mass matrices under the hypothesis that they are 
determined from a minimum principle applied to a generic potential invariant under the $\left[U(3)\right]^5\otimes  {\mathcal O}(3)$ flavor symmetry, 
acting on Standard Model fermions and right-handed neutrinos. Unlike the quark case, we show that hierarchical masses for charged leptons are naturally accompanied by degenerate Majorana neutrinos with one mixing angle close to maximal, a second potentially large, a third one necessarily small, and one maximal relative Majorana phase. The scheme presented here could be tested in the near future via neutrino-less double beta decay and cosmological measurements. }
\section{Introduction}
The Standard Model of particle physics has withstood every experimental check
and has now been confirmed in all of its fundamental aspects. The triumph of the Standard Model
is also the success of the gauge principle as a predictive, powerful and  beautiful
way of describing particle interactions. The recent discovery of the Higgs particle
brings the evidence of a new force of range $1/m_h$ and strength determined by
fermion masses and mixings. As opposed to the three gauge couplings of $\mathcal O (1)$, this force is described by
 at least 13 parameters ranging from $\mathcal O(1)$ to $\mathcal O(10^{-6})$ and encoded in the Yukawa couplings.
An explanation of this structure, explanation which is absent in the Standard Model, will be the answer to what is known as
the flavor puzzle.

The necessary extension of the Standard Model to account for massive neutrinos only adds to this puzzle.
Neutrinos are 6 orders of magnitude lighter than the lightest charged fermion, and the mixing
in the lepton sector is large as opposed to the small angles of the Cabibbo-Kobayashi-Maskawa matrix.

This paper, a summary of the work in Ref.~\cite{Alonso:2013nca}, explores the possibility of the spontaneous breaking of a flavor symmetry as a natural explanation of the observed flavor structure of elementary particles.

\section{The flavor group}
The gauge interactions of the Standard Model (SM) admit a large, global, flavor symmetry. Matter fields in the SM are described by quark and lepton doublets, $q_L$ and $\ell_L$, and by right-handed  singlets corresponding to $up$ and $down$ quarks and to electron-like leptons: $U_R$, $D_R$, $E_R$.  With three quark and lepton generations, 
the flavor group is~\cite{Chivukula:1987py,D'Ambrosio:2002ex}: 

\be 
{\cal G}_0=\left[U(3)\right]^5=U(3)_{q}\otimes U(3)_{U}\otimes U(3)_{D}\otimes U(3)_{{\ell}}\otimes U(3)_{E}\,.
\label{gnumassless}
\ee

Masses for the observed neutrinos can be generated with the see-saw mechanism, by introducing at least two generations of additional Majorana neutrinos, $N_i$. The latter are endowed with a Majorana mass matrix with possibly  
large eigenvalues, and coupled to the lepton doublets by Yukawa interactions.
In analogy with the quark sector, here we assume three Majorana generations. We also assume the maximal flavor symmetry acting on the $N_i$ in the limit of vanishing Yukawa couplings but  non-vanishing Majorana masses, i.e. ${\cal O}(3)$. The flavor group for this case is~\cite{Cirigliano:2005ck}:
\be
{\cal G}=\left[U(3)\right]^5\otimes  {\mathcal O}(3)\,.
\label{gnumass}
\ee
The large flavor group in Eq.~(\ref{gnumass}), of course, does not correspond to observed symmetries. 
In the SM, global symmetries are explicitly broken by the Yukawa couplings of matter fields to the $SU(2)_L$ scalar doublet. Explicitly the Yukawa interactions and neutrino mass terms read:
 \be
- {\cal L}_Y={\bar q}_L {\it Y}_D H D_R +{\bar q}_L {\it Y}_U {\tilde H} U_R+{\bar \ell}_L {\it Y}_E H E_R+{\bar \ell}_L {\it Y}_\nu {\tilde H}  N+\bar N^c \frac{M}{2} N~+ {\rm h.c.},
\label{yukawa}
\ee
where $H$ is the scalar doublet and $\tilde H$ its charge conjugate. Note that the only subgroup of $\mathcal G$ compatible with the above Lagrangian is baryon number, $U(1)_B$, (hypercharge also acts in the Higgs doublet, so its not strictly contained
in $\mathcal G$). For the quarks and charged leptons mass matrices, we find:
\bea
&&M_D= v{\it Y}_D~, \qquad M_U=v{\it Y}_U~, \qquad M_E=v{\it Y}_E \label{mass}~,   \qquad v= \langle 0|H|0 \rangle~. 
\label{vev}
\eea
Integrating over the $N$ fields and keeping the light fields  only, one finds, to lowest order :
\be
\bar N^c\frac{M}{2} N+{\bar \ell}_L {\it Y}_\nu {\tilde H}  N+ {\rm h.c.} \quad \longrightarrow  \quad \frac{1}{2M}{\bar \ell}_L{\it Y}_\nu {\tilde H} {\tilde H}^T {\it Y}_\nu^T {\ell}^C_L+ {\rm h.c.}
\label{yuseesaw}
\ee
which, upon spontaneous breaking of the gauge symmetry, gives the see-saw formula for the light neutrino mass matrix:
\bea
&&m_\nu=\frac{v^2}{M}  {\it Y}_\nu  {\it Y}_\nu^T~.
\label{numass}
\eea
The ${\cal G}$ transformations on ${\it Y}$ that make the Lagrangian formally invariant are as follows:
\bea
&&Y_U \to U_{q} Y_U U_{U}~, \qquad Y_D \to U_{q} Y_D U_{D}~, \label{quarktransf}\\
&& Y_E \to U_{\ell} Y_E U_{E}~, \qquad Y_\nu \to U_{\ell} Y_\nu {\cal O}^T~,
\label{leptransf}
\eea
with the $U$ unitary and ${\cal O}$ real orthogonal, $3\times 3$ matrices. 

The transformation properties of these coupling constants can be understood if they are
somehow the remnants of dynamical objects with actual transformation properties under
the symmetry. This is the option we will discuss in this paper, namely that the Yukawa couplings  are the vacuum expectation values of {\it Yukawa fields}, to be determined by  a minimum principle applied to some potential, $V({\it Y})$, invariant under the full flavor group ${\cal G}$.   In this case, one may use group theoretical methods to identify the {\it natural extrema} and characterize the texture of the resulting Yukawa matrices. 

 The simplest realization of the idea of a dynamical character for the Yukawa couplings is to assume that
 \be
 {\it Y}=\frac{\langle 0|\Phi|0\rangle }{\Lambda}
 \label{vev1}
 \ee
 with $\Lambda$ some high energy scale and $\Phi$ a set of scalar fields with transformation properties such as to make invariant the effective Lagrangians and the potential $V({\it Y})$ under $\cal G$.  
 To avoid the problem of unseen Goldstone bosons,  $ {\cal G}$ may be in fact a local gauge symmetry broken at the scale $\Lambda$, with an appropriate Higgs mechanism, see e.g.~Ref.~\cite{Grinstein:2010ve}.
 
The idea put forward here was considered as early as the sixties by N. Cabibbo, in the attempt to determine theoretically the value of the Cabibbo angle, and group theoretical  methods  were established in Refs.~\cite{Michel:1970mua} and~\cite{Cabibbo:1970rza} to identify the natural extrema of the potential. We review these ideas in the next section.
\section{Natural extrema of an invariant potential} 
We summarize here the elements to identify the {\it natural} extrema of an invariant potential $V(y)$, that is 
those extrema  that are less or not at all dependent from specific tuning of the coefficients in the potential, compared to the generic extrema. 
We do not make any assumption about the convergence of the expansion of the potential in powers of higher-dimensional invariants, as done e.g.~in
Ref.~\cite{Alonso:2012fy,letteraAGHMR}.

The variables $y$ are the field components, transforming as given representations of the invariance group ${\cal G}$. 
In order to be invariant, $V(y)=V[I_i(y)]$, where $I_i$ are the independent invariants one can construct out of $y$. There
are as many independent invariants, $n$, as physical (unaffected by $\mathcal G$ transformations) parameters $y_j$; $i,j=1,..,n$. The crucial point is that the $y$-space has no boundary, while the {\it manifold ${\cal M}$, spanned by $ I_i(y)$, does have boundaries}.

Consider a variation around a given point of the manifold ${\cal M}$, this can be written as:
\be
\delta I_i(y)=\sum_j\frac{\partial I_i}{\partial y_j} \delta y_j\equiv \sum_j J_{ij} \delta y_j \,,
\ee
where $J$ is the Jacobian of the change of ``coordinates". For every point in $y$-space, infinitesimal variations in all $n$ directions are allowed since there is no boundary. In the bulk of the manifold ${\cal M}$, the columns of the Jacobian span a vector basis of dimension $n$
and variations in all directions are also allowed.
 However, for the points of  ${\cal M}$ where the rank of the Jacobian, $r$, is  less than $n$ there exit  $n-r$ directions in $\cal M$ space perpendicular to all columns of the Jacobian. This directions are determined by the linear combinations of rows in $J$ that
 adds to 0. For these points variations in the aforementioned directions are not allowed: we have reached a boundary of dimension $r$.

Boundaries are natural solutions for the minima of a potential since we have that
the extrema of $V(y)$ are to be found by the variational principle:
\be
\delta V=\sum_{ij} \frac{\partial V}{\partial I_i}\frac{\partial I_i}{\partial y_j}\delta y_j=\sum_{ij} \frac{\partial V}{\partial I_i} J_{ij}\delta y_j=0~.
\ee
For arbitrary variations $\delta y_j$ one has a system of $n$ equations. This set of
equations is reduced to $n-r$ for an $r$-dimensional boundary. Note also that the
boundaries are determined from the Jacobian and completely independent of
the potential.

Finally, it can be shown~\cite{Michel:1970mua,Cabibbo:1970rza} that boundaries have associated unbroken subgroups of  $\mathcal G$
of increasing size for decreasing boundary dimension. In connection with the potential minimization, two theorems will be of use in the following:
i) $V$ has always extrema on boundaries having as unbroken subgroup a {\it maximal subgroup} (a subgroup that can be included  only in the full group $\cal G$~\cite{Michel:1970mua}); ii) extrema of $V$ with respect to the points of a given boundary are extrema of $V(y)$~\cite{Cabibbo:1970rza}.

\section{Quarks in three families} 
The counting of parameters in the quark sector goes as follows: 9 complex parameters for each of the
Yukawa matrices, $Y_U$ and $Y_D$, minus the dimension of the group acting on them, $\mbox{dim}(SU(3)^3\times U(1)^2)=26$; that is a total of 10 parameters. Note that baryon number leaves the Yukawa couplings intact. These 10 parameters are no other than the 6 quark masses and 4 mixing parameters in $U_{CKM}$. The invariants can be classified in two types: unmixed invariants,
\be
I_{U^{1}}={\rm Tr}(Y_U Y_U^\dagger)~, \quad   I_{U^{2}}={\rm Tr}[(Y_U Y_U^\dagger)^2]~, \quad I_{U^{3}}={\rm Tr}[(Y_U Y_U^\dagger)^3]~,
\ee
and the same for $Y_D Y_D^\dagger$. The other type comprises the 4 ``mixed" invariants:
\be
\begin{array}{ll}
 I_{U,D}={\rm Tr}( Y_U Y_U^\dagger Y_D Y_D^\dagger)~, \quad 
& I_{U^2,D}={\rm Tr}[(Y_U Y_U^\dagger)^2 Y_D Y_D^\dagger)~,   \\
 I_{U,D^2}={\rm Tr}[Y_U Y_U^\dagger (Y_D Y_D^\dagger)^2]~, \quad 
& I_{(UD)^2}={\rm Tr}[( Y_U Y_U^\dagger Y_D Y_D^\dagger)^2]~. 
\end{array}\label{mixQ}
\ee
Any other invariant can be expressed in terms of these via the Cayley-Hamilton formula~\cite{Jenkins:2009dy}.

Computing the determinant is straightforward and we refer to~\cite{rodrigothesis} for details, but here we will 
use an alternative argument to determine the boundaries.

Unmixed invariants produce extrema corresponding to degenerate or hierarchical patterns as in the chiral case illustrated in Ref.~\cite{Cabibbo:1970rza}.
Mixed invariants involve the CKM matrix $U$, e.g.:
\be\label{qmix}
I_{U,D}={\rm Tr}(Y_U Y_U^\dagger Y_D Y_D^\dagger)\propto \sum_{ij}U_{ij}U^*_{ij}(m^2_U)_i(m^2_D)_j~.
\ee
Extremizing this invariant with respect to the unitary matrix, by the so-called Birkhoff-Von Neumann theorem~\cite{vonneumann}, yields $U_{CKM}$ as a permutation matrix, i.e. a matrix with a $1$ and all other null elements in each row, the $1$ being in different columns. Thus, permutation matrices provide us the singular points on the boundary of the domain, without having to compute the rank of the determinant. The upshot is that, after a relabeling of the $down$ quark coupled to each $up$ quark, we end up with $U_{\rm CKM} =1$.

In the limit of vanishing masses for the first two generations, this solution corresponds to the little  group $U(2)_q \otimes U(2)_U \otimes U(2)_D \otimes U(1)^2$ that is a maximal subgroup of $U(3)_q \otimes U(3)_U \otimes U(3)_D$.

\section{Leptons in three families}
For leptons, we need 15 invariants ($\mbox{dim}(Y_E, Y_\nu)-\mbox{dim}(U(3)^2\times O(3))=36-21$). We may construct unmixed and mixed invariants, as in the quark case. We choose the unmixed ones as:
\be
{\rm Unmixed, E:} \qquad 
I_{E^1}={\rm Tr}( Y_E Y_E^\dagger)~, \quad 
I_{E^2}={\rm Tr}[(Y_E Y_E^\dagger)^2]~, \quad 
I_{E^3}= {\rm Tr}[(Y_E Y_E^\dagger)^3]~,
\label{unmixE}
\ee
and three similar ones  ($I_{\nu^{1-3}}$) using $Y_\nu$. The first type of mixed invariants, completely
analogous to the quark case, are:
\be
{\rm Mixed, ~type ~1}: \qquad 
\begin{array}{ll}
 I_{\nu,E}={\rm Tr}( Y_\nu Y_\nu^\dagger Y_E Y_E^\dagger )~, \quad  
 &  I_{\nu^2,E}={\rm Tr} [( Y_\nu Y_\nu^\dagger)^2 Y_E Y_E^\dagger ]~,  \\
 I_{\nu,E^2}={\rm Tr} [Y_\nu Y_\nu^\dagger (Y_E Y_E^\dagger)^2 ]  ~, \quad 
 & I_{(\nu E)^2}={\rm Tr}[( Y_\nu Y_\nu^\dagger Y_E Y_E^\dagger )^2] ~.
 \end{array}
\label{mixedenu}
\ee
New invariants arise with respect to the quark case as the number of parameters has increased:
\be
{\rm Mixed, ~type ~2}: \qquad 
\begin{array}{ll}
J_{\sigma^1}={\rm Tr}(Y_\nu^\dagger Y_\nu Y_\nu^T  Y_\nu^* )~, \quad 
& J_{\sigma^2}={\rm Tr}[(Y_\nu^\dagger Y_\nu)^2 Y_\nu^T  Y_\nu^* ]~, \\ 
 J_{\sigma^3}={\rm Tr}[(Y_\nu^\dagger Y_\nu Y_\nu^T  Y_\nu^*)^2]~. & 
\end{array}
 \label{LRnu}
 \ee
 Finally, we add two invariants:
 \be
{\rm Mixed, ~type ~3}: \qquad 
\begin{array}{ll}
I_{LR}=\mbox{Tr}\left[{\it Y}_\nu {\it Y}_\nu^T {\it Y}_\nu^* {\it Y}_\nu^\dagger {\it Y}_E {\it Y}_E^\dagger \right]~, \\
I_{RL}=\mbox{Tr}\left[{\it Y}_\nu {\it Y}_\nu^T{\it Y}_E^* {\it Y}_E^T  {\it Y}_\nu^* {\it Y}_\nu^\dagger {\it Y}_E {\it Y}_E^\dagger \right]~.
\end{array}
\qquad\qquad 
\ee
Let us introduce the bi-unitary parametrization for the neutrino Yukawa, $Y_\nu=U_L \mathbf y U_R$, with $\mathbf y=\mbox{diag} (y_1,y_2,y_3)$ and $U_{L,R}$ unitary. The impact of the mixed operators can then be simplified and discussed in terms of the different types.

Type 1 invariants depend on $U_L$ but not $U_R$, and the minimization of the former yields a permutation matrix in analogy with $U_{CKM}$ but with the important difference that $U_L$ is {\it not} the lepton  mixing matrix. Type 2 invariants conversely only depend on $U_R$; for example, invariant $J_{\sigma^1}$ reads:
\be
J_{\sigma^1}={\rm Tr}(Y_\nu^\dagger Y_\nu Y_\nu^T  Y_\nu^* )= \sum_{ij}(U_R U_R^T)_{ij}(U_R^*U_R^\dagger)_{ij}y^2_i \,y^2_j.
\ee
Direct comparison with Eq.~(\ref{qmix}) reveals that $U_R U_R^T$ is now a permutation matrix when extremized. 
To extract the consequences of this result we shall look at the neutrino mass matrix.
First, we use the freedom in the neutrino labeling to set $U_L=1$ in the basis where charged leptons are ordered according to: 
${\it Y}_E=\mbox{diag}\,(y_e, y_\mu,y_\tau).$
Using the expression in Eq.(\ref{numass}), assuming degenerate eigenvalues for $Y_\nu$ for reasons given below, and
taking one of the possible permutations for $U_R U_R^T$ leads to:
\be\label{numass2}
m_\nu=\frac{v^2}{M}~Y_\nu   Y_\nu^T=\frac{v^2}{M}~\mathbf{y}\,U_R U_R^T\,\mathbf{y}=\frac{y^2v^2}{M}~\left(\begin{array}{ccc}1 & 0 & 0\\ 0 & 0 & 1  \\0 &1  & 0 \end{array}\right)~.
\label{solution}
\ee
From the second identity in Eq.~(\ref{numass2}) we find that the absolute values of neutrino masses are degenerate
and equal to $y^2v^2/M$ whereas the mixing matrix is:
\bea
&&U_{\rm PMNS}^{(0)}=\left(\begin{array}{ccc}1 & 0 & 0\\ 0 & 1/\sqrt{2} & 1/\sqrt{2}\\ 0 & -1/\sqrt{2} & 1/\sqrt{2} \end{array}\right)~, 
\qquad \Omega= {\rm diag}(1,1,i)~,
\label{pmns0}
\eea
where $\Omega$ is the diagonal matrix of Majorana phases. One may fear that the
degeneracy of neutrino masses makes the mixing matrix unphysical. This is certainly
true at this stage for the first two mass eigenstates, but not for their mixing with the
third; there is a relative maximal Majorana phase between them that makes them physically distinct.
The maximal angle appearing in Eq.~(\ref{pmns0}) is therefore physical and can be taken
as the atmospheric angle, experimentally determined to be close to maximal.

This striking difference with quarks arose in spite of treating quark and leptons 
in the same symmetry footing and it is a promising starting point. 
From an algebraic point of view, one can say that this method predicts at first
order that the quark mixing matrix is a permutation matrix whereas the lepton mixing
matrix is the ``square root" of a permutation matrix.

The choice of degeneracy for the diagonal entries in $Y_\nu$ and therefore neutrino masses is not
chosen here for simplicity but
for necessity. For arbitrary entries in $\mathbf y$, the first two neutrino eigenstates are not degenerate; this
would make their relative angle in Eq.~(\ref{pmns0}) physical and equal to $0$. The angle in the
$1$-$2$ sector is the solar angle and it is very far from vanishing.

We are therefore forced to degenerate neutrinos if we want to explain the mixing pattern.
The reason why one can do this is that, at first order for degenerate eigenstates, the solar angle is unphysical and 
a ``flat" direction. Perturbations in the neutrino mass matrix can then introduce both a split in neutrino masses 
and a large solar angle. Let us show this explicitly with selected perturbations in the neutrino mass matrix, for
the general scenario we refer to the original work~\cite{Alonso:2013nca}: 
\be
m_\nu = \frac{v^2 y}{M}\left(\begin{array}{ccc}1 +\sigma & \epsilon & \epsilon
\\\epsilon & 0 & 1\\\epsilon & 1& 0 \end{array}\right)~,
\ee
with $\epsilon,\sigma\ll1$ and real. 
A simple calculation leads to
\bea
&& U_{\rm PMNS}=\left[ \begin{array}{ccc}
{\cos\theta_{12}}  & -\sin\theta_{12} & 0 \\[2mm]
               \displaystyle{\frac{\sin\theta_{12}}{\sqrt{2}}}\
           &  \displaystyle{\frac{\cos\theta_{12}}{\sqrt{2}}} \
           & -\displaystyle\frac{1}{\sqrt{2}} \\[3mm]
               \displaystyle{\frac{\sin\theta_{12}}{\sqrt{2}}}\
           &  \displaystyle{\frac{\cos\theta_{12}}{\sqrt{2}}}\
           & +\displaystyle\frac{1}{\sqrt{2}} \end{array}\right]~,\quad\tan(2\theta_{12})=2\sqrt{2}\epsilon/\sigma\,,
\eea
and the induced mass splinting between the first two eigenstates is $2\sqrt{2}y^2v^2\epsilon/(\sin(2\theta_{12})M)$. A large $\theta_{12}$ follows from its expression as a ratio of perturbations which need not be small.

For general perturbations the PMNS matrix features a generically large $\theta_{12}$ (that we cannot compute in absence of firm predictions for the values of 
$ \epsilon$ and $\sigma$, but that does not goes to zero in the limit of vanishing perturbations), 
$\theta_{23}$ close to $\pi/4$, and $\theta_{13}$ generically small. 
The spectrum is almost degenerate, with normal or inverted hierarchy according to the signs of the perturbations, and 
mass splittings not correlated to the mixing matrix. Nonetheless, assuming that the perturbations that cause the mass splitting
are of the same order as those generating $\theta_{13}$, one can estimate a lightest neutrino mass of $0.1$ eV.
This size is within reach of the next generation of $0\nu\nu$ double beta decay experiments~\cite{pdg},  and possibly of
cosmological measurements \cite{Ade:2013zuv}.
Note also that the size of the perturbations is not far from what could be deduced from the charged lepton spectrum, 
treating $m_\mu/m_\tau \approx 0.06$ as estimate of the sub-leading terms.

\section{Conclusions and outlook} 
We have assumed that the structure of quark and lepton mass matrices derives
from a minimum principle, with the maximal flavor symmetry $\left[U(3)\right]^5\otimes  {\mathcal O}(3)$
 and a minimal breaking due to the vevs of fields 
transforming like the Yukawa couplings.  For leptons we find a natural solution correlating large mixing angles and degenerate neutrinos. This solution 
generalizes to three familes and arbitrary invariant potential the results found in Ref.~\cite{Alonso:2012fy,letteraAGHMR}.
Subject to small perturbations, the solution can reproduce the observed pattern of neutrino masses 
and mixing angles. Our considerations lead to a value of the common neutrino mass 
that is within reach of the next generation of neutrinoless double beta decay experiments.

\section*{Acknowledgements}
The author would like to thank the organizers of the conference for their kind invitation.

\end{document}